\documentstyle[12pt]{article}

\begin{document}
\baselineskip 0.7cm

\setcounter{footnote}{1}

\begin{titlepage}


\vskip 0.35cm
\begin{center}
{\large \bf
Duality of a Supersymmetric Standard Model
}
\vskip 1.2cm
Nobuhiro Maekawa
\footnote
{e-mail: maekawa@gauge.scphys.kyoto-u.ac.jp}

\vskip 0.4cm

{\it Department of Physics, Kyoto University,\\
      Kyoto 606-01, Japan}

\vskip 1.5cm

\abstract{
We examine a dual theory of a Supersymmetric Standard Model(SSM) in
terms of an $SU(3)_C$ gauge group. In this scenario,  it is naturally
understood that at least one quark (the top quark) should be heavy, i.e.,
almost the same
order as the weak scale.
Moreover,  the supersymmetric Higgs mass parameter $\mu$
can naturally be expected to be small. This model automatically induces
nine pairs of composite Higgs fields,
which may be observed in the near future.
}

\vskip 1.5cm

\end{center}
\end{titlepage}

%
%
%
%
Recently, it has become clear that certain aspects of four dimensional
supersymmetric field theories can be analyzed exactly
\cite{duality,holomorphy,both,DSB}.
One of the most interesting aspects is ``duality''
\cite{duality}.
By using
``duality'', we can infer the low energy effective theory of a strong
coupling gauge theory. Does nature use this ``duality''? In this
paper, we would like to discuss a duality of a Supersymmetric Standard
Model(SSM).

First we would like to review Seiberg's duality.
Following his discussion
\cite{duality}, we examine $SU(N_C)$ supersymmetric (SUSY) QCD with
$N_F$ flavors of
chiral superfields,
\begin{center}
\begin{tabular}{|l|c|c|c|c|c|} \hline \hline
       & $SU(N_C)$ & $SU(N_F)_L$ & $SU(N_F)_R$ & $U(1)_B$ & $U(1)_R$ \\ \hline
$Q^i$    & $N_C$ & $ N_F$ &  1  &   1    & $(N_F-N_C)/N_F$  \\
$\bar Q_j$ & $\bar N_C$ & 1 & $\bar N_F$ & $-1$ & $(N_F-N_C)/N_F$ \\
\hline \hline
\end{tabular}
\end{center}
which has the global symmetry $SU(N_F)_L\times SU(N_F)_R \times U(1)_B
\times U(1)_R$. This theory is called the electric theory.
In the following, we would like to take $N_F\geq N_C+2$, though in the
case $N_F\leq N_C+1$ there are a lot of interesting features
\cite{both,ADS,AKMRV,NSVZ}.
Seiberg suggests
\cite{duality}
 that in the case $N_F\geq N_C+2$ at the low energy scale
the above theory is
equivalent to the following $SU(\tilde N_C)$ SUSY QCD theory $(\tilde
N_C=N_F-N_C)$ with $N_F$ flavors of chiral superfields $q_i$ and $\bar
q^j$ and meson fields $T^i_j$,
\begin{center}
\begin{tabular}{|l|c|c|c|c|c|} \hline \hline
       & $SU(\tilde N_C)$ & $SU(N_F)_L$ & $SU(N_F)_R$ & $U(1)_B$ &
       $U(1)_R$ \\ \hline
$q_i$    & $\tilde N_C$ & $ \bar N_F$ &  1  &  $N_C/(N_F-N_C)$ & $N_C/N_F$  \\
$\bar q^j$ & $\bar {\tilde N_C}$ & 1 & $ N_F$ & $-N_C/(N_F-N_C)$ & $N_C/N_F$ \\
$T^i_j$ & 1 & $N_F$ & $\bar N_F$ & 0 & $2(N_F-N_C)/N_F$ \\
\hline \hline
\end{tabular}
\end{center}
and with a superpotential
\begin{eqnarray}
W=q_iT^i_j\bar q^j.
\end{eqnarray}
This theory is called the magnetic theory.
The above two theories satisfy the 't Hooft
anomaly matching conditions
\cite{tHooft}.
Moreover Seiberg showed that they are
consistent with the decoupling theorem
\cite{decoupling}.
Namely, if we introduce a mass
term only for superfields $Q^{N_F}$ and $\bar Q_{N_F}$
\begin{eqnarray}
W=m Q^{N_F}\bar Q_{N_F}
\end{eqnarray}
in the original (electric) theory, the dual (magnetic)
theory has vacuum expectation values(VEVs)
$\left<q\right>=\left<\overline{q}\right>=\sqrt{m}$
and $SU(N_f-N_c)$ is broken to $SU(N_f-N_c-1)$,
which is consistent with
the decoupling of the heavy quark in the original theory.

Next, we would like to discuss a duality of a SUSY Standard
Model (SSM). We introduce ordinary matter superfields
\begin{eqnarray}
&&Q^i_L=(U_L^i, D_L^i):(3,2)_{1\over 6},\quad U_{Ri}^{c}:(\bar
3,1)_{-{2\over 3}},\quad
D_{Ri}^{c}:(\bar 3,1)_{1\over 3} \nonumber \\
&&L^i=(N_L^i,E_L^i):(1,2)_{-{1\over 2}},\quad
E_{Ri}^{c}:(1,1)_1,\quad\quad i=1,2,3,
\end{eqnarray}
which transform under the gauge group $SU(3)_{\tilde C}\times
SU(2)_L\times U(1)_Y$. There are no Higgs superfields.
We would like to examine the magnetic theory of this electric theory
with respect
to the gauge group
$SU(3)_{\tilde C}$. In the following, we neglect the lepton
sector for simplicity. Since $N_F=6$, the dual gauge
group is also $SU(3)_C$ ( $\tilde N_C=N_F-N_C$ ), which we would like
to assign to the QCD gauge group.
A subgroup, $SU(2)_L\times U(1)_Y$, of
the global symmetry group $SU(6)_L\times
SU(6)_R\times U(1)_B\times U(1)_R$ is  gauged.
When we assign $Q=(U_L^1,D_L^1,U_L^2,D_L^2,U_L^3,D_L^3)$ and $\bar
Q=(U_R^{c1},D_R^{c1},U_R^{c2},D_R^{c2},U_R^{c3},D_R^{c3})$, the
$SU(2)_L$ generators are given by
\begin{eqnarray}
I_L^a=I_{L1}^a+I_{L2}^a+I_{L3}^a,\quad a=1,2,3,
\end{eqnarray}
where $I_{Li}^a$ are generators of $SU(2)_{Li}$ symmetries which
rotate $(U_L^i, D_L^i)$,
and the generator of hypercharge $Y$ is given by
\begin{eqnarray}
Y={1\over 6}B-(I_{R1}^3+I_{R2}^3+I_{R3}^3),
\end{eqnarray}
where $I_{Ri}^a$ are generators of $SU(2)_{Ri}$ symmetries which
rotate $(U_{Ri}^c, D_{Ri}^c)$.
In this theory, the global symmetry group is $SU(3)_{QL}\times
SU(3)_{UR}\times SU(3)_{DR}\times U(1)_B\times U(1)_R$.
Then we can write down the quantum numbers of dual fields;
\begin{eqnarray}
&&q_{Li}=(d_{Li}, -u_{Li}):(3, \bar 2)_{1\over 6},\quad u_R^{ci}:(\bar
3,1)_{-{2\over 3}},\quad
d_R^{ci}:(\bar 3,1)_{1\over 3} \nonumber \\
&&M^i_j:(1,2)_{-{1\over 2}},\quad N^i_j:(1,2)_{1\over 2}
\end{eqnarray}
under the standard gauge group $SU(3)_C\times SU(2)_L\times U(1)_Y$.
Here $M^i_j\sim Q_L^iU_{Rj}^c$ and $N^i_j\sim Q_L^iD_{Rj}^c$ are the
meson fields and we assign
$q=(d_L^1, -u_L^1, d_L^2, -u_L^2, d_L^3, -u_L^3)$ and $\bar
q=(d_{R}^{c1}, -u_{R}^{c1}, d_{R}^{c2}, -u_{R}^{c2}, d_{R}^{c3}, -u_R^{c3})$.
It is interesting that the matter contents of both theories are almost
the
same. The difference is the existence of nine pairs of Higgs
superfields $M^i_j$ and $N^i_j$
and their Yukawa terms coupling to ordinary matter superfields,
\begin{eqnarray}
W=-y_u q_L^i N_i^j u_{Rj}^c+y_d q_L^iM_i^jd_{Rj}^c.
\end{eqnarray}
Here neglecting the effect of $U(1)_Y$ gives $y_u=y_d$. The Yukawa
couplings can be expected to be of order one because of the strong
dynamics
\footnote{
By using the superconformal algebra ($D=3|R|/2$),
the conformal dimension $D$ of the meson fields can be calculated as 3/2,
which is
much different from the conformal dimension of free fields. Therefore
the meson fields must have some strong interaction. It is natural to
regard the Yukawa interaction as the strong one.
Though this argument is reliable only at the infra-red fixed
point, we expect that the Yukawa  couplings remains large even if the
coupling is not at the infra-red fixed point.}
{}.

In the following, we only assume that the duality can be realized
even if SUSY breaking terms
\begin{eqnarray}
{\cal L}_{SB}^e=\sum_{i=1}^3 \left( m_{Qi}^2
  |Q^i|^2+m_{Ui}^2|U_i|^2+m_{Di}^2|D_i|^2\right)+{1\over
  2}\sum_{a=1,2,3} \mu_a \lambda_a\lambda_a,
\end{eqnarray}
where $\lambda_a$ are gauginos,
are introduced and that the scale $\Lambda$ at which the duality
becomes a bad
description is higher than the SUSY breaking scale. We will discuss
these assumptions later. In general we
can take $m_1^2\ge m_2^2\ge m_3^2$.
The global symmetry is broken to $U(1)^8$
by these SUSY breaking terms.
By using the perturbation
\cite{Peskin}
we can get the SUSY breaking terms of the dual theory
\begin{eqnarray}
{\cal L}_{SB}^m&=&\sum_{i=1}^3 \left( m_{qi}^2
  |q_i|^2+m_{ui}^2|u^i|^2+m_{di}^2|d^i|^2\right)+\sum_{i,j=1}^3
\left( m_{Mij}^2 |M_i^j|^2+m_{Nij}^2 |N_i^j|^2\right) \nonumber \\
&&+{1\over 2}\sum_{a=1,2,3} \mu_a \lambda_a\lambda_a , \\
 m_{qi}^2&\sim& m_{Qi}^2,\quad m_{ui}^2\sim m_{Di}^2,\quad m_{di}^2\sim
 m_{Ui}^2,\quad m_{Mij}^2\sim m_{Qi}^2+m_{Uj}^2,\quad m_{Nij}^2 \sim
m_{Qi}^2+m_{Dj}^2. \nonumber
\end{eqnarray}
The Higgs can have a vacuum expectation value(VEV) radiatively
\cite{Inoue}.
By using the renormalization group equations
\begin{eqnarray}
{d\over dt}m_{Nij}^2&=&{1\over 8 \pi^2}\left( \tilde N_Cy_u^2(m_{Nij}^2
 +m_{qi}^2+m_{uj}^2) + {1\over 2} g_1^2{\rm
   tr}(Ym^2)-3g_2^2\mu_2^2-g_1^2\mu_1^2 \right),\\
{d\over dt}m_{Mij}^2&=&{1\over 8 \pi^2}\left( \tilde N_Cy_d^2(m_{Mij}^2
 +m_{qi}^2+m_{dj}^2) -{1\over 2}g_1^2{\rm
   tr}(Ym^2)-3g_2^2\mu_2^2-g_1^2\mu_1^2 \right),\\
{d\over dt}m_{qi}^2&=&{1\over 8 \pi^2}\left( 3(y_u^2+y_d^2)m_{qi}^2
 +\sum_j^3(y_u^2(m_{Nij}^2+m_{uj}^2)+y_d^2(m_{Mij}^2+m_{dj}^2))
\right.\nonumber \\
&&\left.+{1\over 6}g_1^2{\rm tr}(Ym^2)
-{16\over 3} g_3^2 \mu_3^2-3g_2^2\mu_2^2
 -{1\over 9}g_1^2\mu_1^2 \right),\\
{d\over dt}m_{ui}^2&=&{1\over 8 \pi^2}\left( 3y_u^2m_{ui}^2
 +\sum_j^3y_u^2(m_{Nij}^2+m_{uj}^2)
-{2\over 3}g_1^2{\rm tr}(Ym^2)-{16\over 3} g_3^2 \mu_3^2
 -{16\over 9}g_1^2\mu_1^2 \right),\\
{d\over dt}m_{di}^2&=&{1\over 8 \pi^2}\left( 3y_d^2m_{di}^2
 +\sum_j^3y_d^2(m_{Mij}^2+m_{dj}^2)
+{1\over 3}g_1^2{\rm tr}(Ym^2)-{16\over 3} g_3^2 \mu_3^2
 -{4\over 9}g_1^2\mu_1^2 \right),
\end{eqnarray}
we can expect that the smallest mass term at the low energy scale will
be $m_{M33}^2$ or $m_{N33}^2$ unless $\mu_a^2<<m^2$.
The tree Higgs potential is
\begin{eqnarray}
V_{H}&=&\sum_{ij}m_{Nij}^2|N_i^j|^2+m_{Mij}^2|M_i^j|^2
+{1\over 8}g_2^2\sum_a |\sum_{ij}(N_j^{\dagger i} \tau_a N_i^j
+M_j^{\dagger i} \tau_a M_i^j)|^2\nonumber \\
&&+{1\over 8}g_1^2|\sum_{ij}(N_j^{\dagger i} N_i^j-M_j^{\dagger i}
M_i^j)|^2.
\end{eqnarray}
If only $m_{Nij}^2$ is negative, we can take  $\langle
N_{33}\rangle=(v,0)$.
In this case,
from the Higgs potential (15) we can find that $N_i^j$ have a
tendency to have a VEV with an unbroken electromagnetic
interaction $U(1)_{EM}$, on the other hand $M_i^j$ have a tendency to
have a VEV breaking the $U(1)_{EM}$. In order to avoid breaking the
$U(1)_{EM}$, we introduce the
following conditions
\begin{eqnarray}
m_{Ui}^2-m_{D3}^2>m_Z^2,\quad i=1,2,3,
\end{eqnarray}
where $m_Z$ is the mass of the $Z$ boson. Moreover if we introduce the
conditions
\begin{eqnarray}
m_{Di}^2-m_{D3}^2&>&m_Z^2,\\
m_{Qi}^2-m_{Q3}^2&>&m_Z^2,\quad i=1,2,
\end{eqnarray}
only $N_{33}$ can have a VEV. By the above discussions, we will not
claim that only the top quark is naturally heavy. We would like to
emphasize that
the large Yukawa couplings are naturally understood and it is possible
that only the top quark has a large mass by imposing some conditions
which are not so unnatural.

It is also interesting that the global symmetry $SU(3)_Q\times
SU(3)_{UR}\times SU(3)_{DR}$ forbids the term $\mu MN$. Namely, we can
understand the smallness of the SUSY Higgs mass
$\mu$. Phenomenologically the mass $\mu$ should be around the SUSY
breaking scale $m_{SB}$. We do not have a definite explanation for realizing
$\mu\sim m_{SB}$.
However, we would like to emphasize here that we can
naturally introduce the global symmetry which forbids the mass term.


Of course, in this model, if further Higgs fields have some VEVs,
massless Nambu-Goldstone bosons appear, because the global $U(1)$
symmetries still exist. Therefore in order to give masses to the other
quarks without massless Nambu-Goldstone bosons, we must break the
global $U(1)$ symmetries explicitly. There are some ways to break the
symmetries. One possibility is to introduce the Higgs fields and the
Yukawa couplings in the electric theory,
which can break the global
symmetries.
As another possibility, if we introduce vector-like fields,
we can break the
symmetries by the SUSY breaking terms
\cite{Takahashi}.
Even when we do not introduce new fields,
we can break all
the global $U(1)$ symmetries except $U(1)_B$ by introduce SUSY
breaking terms violating R-parity
\cite{Joe}.


It is a serious problem that the leptons are massless in
this model. There are several mechanisms which can induce the Yukawa
couplings of leptons
\cite{Takahashi,Joe}.
One possibility is to introduce the Pati-Salam gauge group
\cite{Takahashi},
and
another one is to break the R-parity
\cite{Joe}. Here we do not discuss these possibilities further.

Finally, we would like to discuss the assumptions introduced
previously. Does duality hold even with SUSY breaking terms?
This is still an open question though some people
\cite{Peskin,Evans}
analyze this subject.
Since some useful techniques such as holomorphy cannot be used in the
analysis with SUSY breaking terms, it is difficult to get a definite
answer. Therefore in this paper we only assume that there is a duality
in this case.

Where is the scale at which the dual description is broken?
Since Seiberg's discussion using the superconformal algebra
is effective only in the conformal phase, the duality might exist only
in the conformal phase (or close to it).
If the duality is effective only in the conformal phase, our
model would be
invalid because the QCD coupling ($\alpha_s \sim 0.11-0.12$) may be
too far from the infra-red fixed coupling ($\alpha^* \sim 0.6$).
However, since the discussions
about 'tHooft anomaly matching conditions and about the consistency
with the decoupling are effective even in the non-conformal phase,
it is not strange that the duality exists even with the coupling far
from the infra-red fixed point.
If the both electric and magnetic theories are asymptotically free, the
dual description will be broken at the scale at which the gauge
coupling is not so far from the infra-red fixed point. This is because
the difference will be obvious when the couplings of both theories are
small enough to use the perturbations
\cite{Aharony}.
However, what we would like to emphasize here is the possibility
that one of them is not asymptotically free. Since both theories with
$3N_C/2<N_F<3N_C$ have
infra-red fixed points, there is the asymptotically non-free
phase. Namely, when the gauge coupling is above the infra-red
fixed coupling, the theory is in the asymptotically non-free phase.
If this possibility is correct, the dual description may be realized
even with the coupling far from the infra-red fixed point, because
both theories cannot have the small couplings
simultaneously.
In this model, if the electric theory is asymptotically non-free, then the
smallness of the QCD coupling in magnetic theory may be realized.
Of course we know that the duality is realized only in the low energy
scale, because some other massive states will exist ({\it e.g.}, in the
dual theory of the dual theory, meson fields become massive.).
However you should notice that the mass scale of the
heavy particles which should be decoupled is unknown.

Does the above scenario work even if the QCD scale is lower than the
SUSY breaking scale $m_{SB}$?
Here we will only note that if such a duality persists even with SUSY
breaking $m_{SB}$,
the QCD scale, which is defined by the divergence of the gauge
coupling, should be smaller than the SUSY breaking scale $m_{SB}$. The
reason is the following. Since the SUSY models have
an infra-red fixed point, the gauge coupling does not diverge
without SUSY breaking. Under the SUSY breaking scale, the running of
the coupling changes and the coupling can diverge at some scale,
$\Lambda_{QCD}< m_{SB}$.



In summary, ``duality'' may be interesting even in the real world.
We applied this technique to a Supersymmetric Standard Model. Though
this model is not a complete model phenomenologically, there are some
interesting features. The Higgs fields are induced as composites,
a heavy top quark is naturally understood, and there is a global
symmetry which forbids the SUSY Higgs mass terms.
The model also predicts nine pairs of Higgs superfields.
Since we do not know which theories is connected to the real high
energy theory, we believe that the duality gives
us very rich possibilities for model building.

We are grateful to M. Peskin for his lectures on recent developments of
the SUSY QCD at the Ontake Summer
Institute.
We would like to thank T. Kawano, J. Sato, and M. Strassler for
useful discussions.
We also thank A. Bordner for proofreading this manuscript.

\end{document}